\documentclass[superscriptaddress,amsmath,amssymb,reprint,showkeys,showpacs,aps,prl]{revtex4-2}
\usepackage{graphicx} 

\usepackage{color}

\newcommand{\ket}[1]{\left| #1 \right\rangle}
\newcommand{\bra}[1]{\left\langle #1 \right|}
\newcommand{\braket}[2]{\left\langle #1 | #2 \right\rangle}
\newcommand{\ketbra}[2]{\left| #1 \right\rangle \!\!\left\langle #2 \right|}

\date{\today}
\preprint{APS/123-QED}
\usepackage{hyperref}

\begin{document}
\title{Continuous-Variable Quantum State Tomography Enabled by Quantum Mirrors}

\author{M. Uria}
\email{maruria@udec.cl}
\affiliation{Facultad de Ciencias F\'isicas y Matem\'aticas, Departamento de F\'isica, Universidad de Concepci\'on, Concepci\'on, Chile}

\author{A. Moya}
\email{amarumoya@udec.cl}
\affiliation{Facultad de Ciencias F\'isicas y Matem\'aticas, Departamento de F\'isica, Universidad de Concepci\'on, Concepci\'on, Chile}
\affiliation{ANID - Millennium Science Iniciative Program - Millennium Institute for Research in Optics}

\author{C. Hermann-Avigliano}
\email{carla.hermann@uchile.cl}
\affiliation{ANID - Millennium Science Iniciative Program - Millennium Institute for Research in Optics}
\affiliation{Departamento de F\'isica, Facultad de Ciencias F\'isicas y Matem\'aticas, Universidad de Chile, Santiago, Chile}

\author{P. Solano}
\email{psolano@udec.cl}
\affiliation{Facultad de Ciencias F\'isicas y Matem\'aticas, Departamento de F\'isica, Universidad de Concepci\'on, Concepci\'on, Chile}

\author{A. Delgado}
\email{aldelgado@udec.cl}
\affiliation{Facultad de Ciencias F\'isicas y Matem\'aticas, Departamento de F\'isica, Universidad de Concepci\'on, Concepci\'on, Chile}
\affiliation{ANID - Millennium Science Iniciative Program - Millennium Institute for Research in Optics}

\begin{abstract}
In quantum technologies, continuous-variable systems offer advantages over their discrete counterparts. However, continuous-variable tomography suffers from exponentially growing sample complexity. We propose protocols using \textit{quantum mirrors} to transfer the complete information of incident photonic states onto a control atomic system. This enables full photonic state characterization through measurements on the control atom alone, realized via kernel functions, direct wavefunction reconstruction, and pointwise Wigner function measurements. Our approach overcomes the limitations of conventional photon counting, statistical inference, and inverse transformation, providing a robust framework for benchmarking and verifying non-Gaussian states in continuous-variable quantum optics.
\end{abstract}

\maketitle

{\it Introduction.---} 
The development of quantum technologies \cite{wang_2025} such as computing \cite{PsiQuantum_2025, Madsen_2022, Zhong_2020}, communication \cite{Mele_2025}, and sensing \cite{Aasi_2023} requires benchmarking, verification, and optimization.
These tasks can be implemented by accurately characterizing quantum states via quantum state tomography (QST) \cite{Lvovsky_2009}.
Although these methods have been successfully implemented in experimental settings \cite{Banaszek_Radzewicz_1999, Kirchmair_2013,Wang_2009,Vlastakis_2013,Bertet_2002,Haroche_2008,Hofheinz_2009,Wang_2016,Landon-Cardinal_2018, laiho_2010}, they are resource-intensive and computationally demanding in terms of post-processing.

A particularly challenging scenario for QST appears in continuous variable (CV) systems, as they show a significantly higher sample complexity, which scales exponentially with the energy of the state \cite{MeleandMele_2025},   in contrast to their discrete counterparts \cite{ODonnell_2016,Haah_2017}. Furthermore, QST via quasiprobability distribution functions requires prior knowledge of the state and demanding mathematical tools, such as inverse linear transformations or maximum-likelihood estimations \cite{Lvovsky_2009,Leonhardt_1997,Vogel1989, strandberg_2022, ahmed_sanchez_munoz_nori_kockum_2021,Lutterbach_Davidovich_1997, shen_heeres_reinhold_jiang_liu_schoelkopf_jiang_2016,Boas_2001, solano_2005, leibfried_1996, Lundeen2011}. Given the advantages provided by CV systems, including their natural compatibility with scalable photonic and hybrid architectures \cite{andersen_2010,andersen_2015}, together with their applications in quantum sensing \cite{caves_1981,escher_davidovich_2011, polino_valeri_2020} and hardware-efficient quantum error correction \cite{gottesman_2001,guillaud_mirrahimi_2019}, the development of reliable CV-QST is essential.

\begin{figure}[t]
    \centering
    \hspace{-0.25cm}\includegraphics[width=\columnwidth]{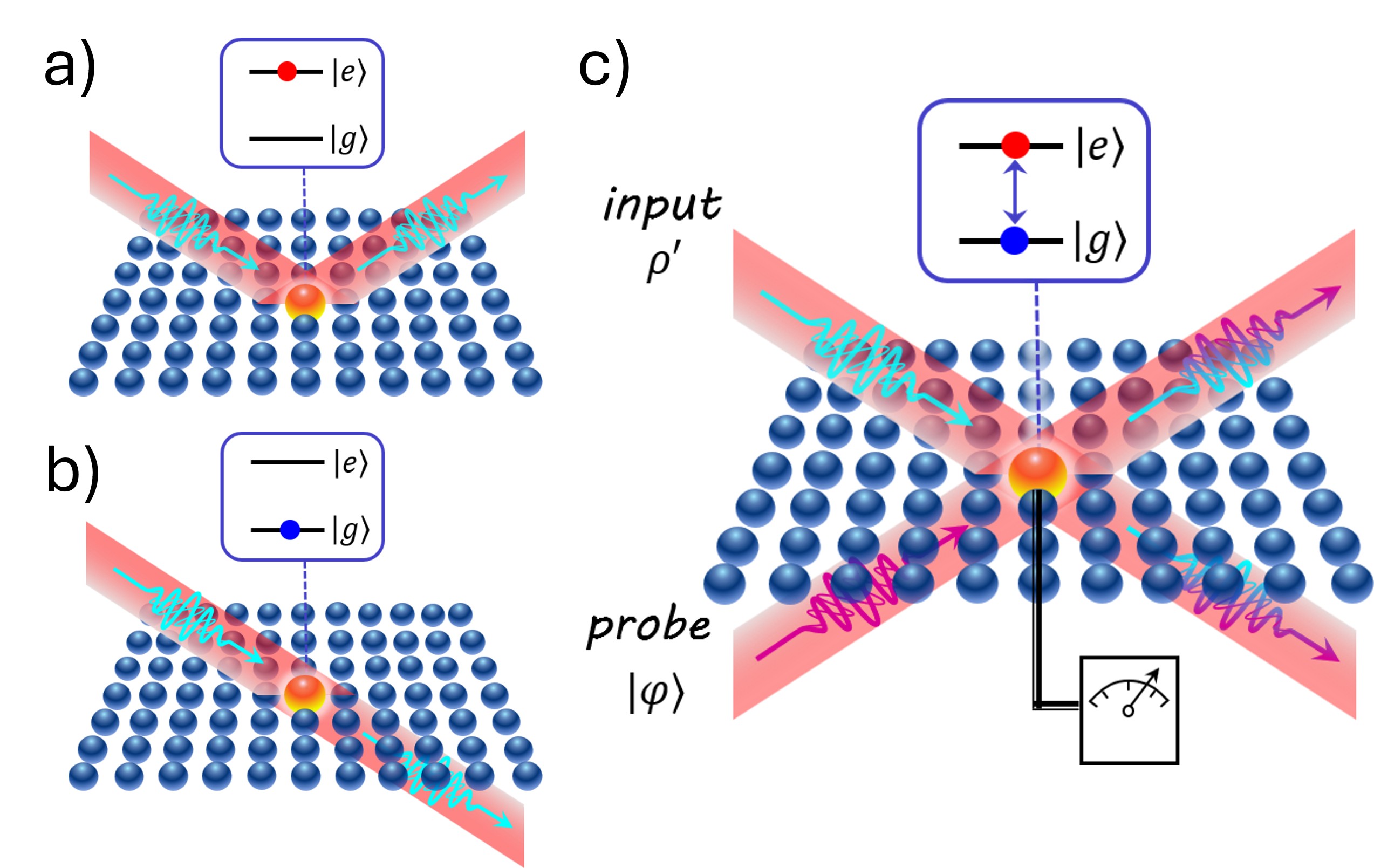}
    \caption{
    (a) The control atom in state $\ket{e}$ changes the optical response of the bulk, which generates a complete reflection, thus ''gaining'' a phase shift. (b) The control atom in state $\ket{g}$ generates complete transmission of the modes going through the mirror. (c) Representation of the quantum mirror in a superposition state and its effect on the input photonic modes.
    }
    \label{fig:QM}
\end{figure}

Addressing these challenges requires novel platforms and protocols for the efficient characterization of CV systems \cite{gorlach_2024, jeon_kang_choi_kim_you_kim_2025,fluhmann_home_2020,bazavan__ballance_srinivas_2026, krisnanda_fontaine_copetudo_song_lee_huang_valadares_liew_gao_2025}. Atomic-scale materials, such as quantum metasurfaces, are currently being explored for quantum-state engineering and hold promise as platforms for quantum optics \cite{Bekenstein2020,Shahmoon2017, Srakaew2023, RuiWeiRubio2020}, thus motivating their potential application in CV-QST.

Here, we introduce a set of protocols for CV-QST that are capable of obtaining the wavefunction and all quasiprobability distributions without a priori information and, in certain cases, even without mathematical transformations for their reconstruction. Our proposal is based on \textit{quantum mirror} devices-- optical interfaces with switchable reflectivity controlled by the quantum state of an ancillary atom \cite{Bekenstein2020}-- where two incident photonic modes can entangle with the control atom upon interaction, as illustrated in Fig.~\ref{fig:QM}. Given the knowledge of one of the modes, we propose protocols to reconstruct arbitrary unknown CV states by performing measurements exclusively on the control atom. These protocols present the first proposal for directly measuring the value of the quasiprobability distributions for propagating electromagnetic modes at any point in phase space, which can be used to monitor nonclassicality. The protocols also bypass some of the limitations of homodyne detection, such as photodetection inefficiency, mode mismatch, bandwidth limitations, and electronic and excess noise, by transferring these sources of errors to those associated with two-level atom tomography. The platform and protocols introduced here broaden the toolbox of CV-based quantum technologies.

{\it Quantum Mirrors.---} Atomic ensembles coupled to a control two-level atom, whose state induces perfect reflection or complete transparency for incoming electromagnetic modes \cite{Bekenstein2020}, are usually referred to as quantum metasurfaces or {\it quantum mirrors} (QM). Coherent light-matter interaction enables the generation of entanglement between electromagnetic modes and the control two-level atom. These devices exhibit complex transmission and reflection coefficients \cite{Campos1989}, $t$ and $r$, related by 
\begin{equation}
t = e^{i\phi} + r, \label{eq:tandr}
\end{equation}
where $\phi$ is the scattering phase acquired by each mode upon interaction. Although $\phi$ is often considered a global phase that can be neglected, it plays a central role in quantum mirrors by defining the relative phase in the quantum superposition between the transmitted and reflected fields (see Supplementary Material \cite{SM}).

The action of a mirror over two arbitrary incoming photonic states $\{\ket{\psi},\ket{\varphi}\}$ is given by $\hat{U}_M$, under appropriate parameterization (see Supplementary Material \cite{SM})
\begin{equation}
    \hat{U}_{M} \ket{\psi}_0 \ket{\varphi}_1 = e^{-i \phi \hat{n}_0} \hat{\Pi}_0 \ket{\varphi}_{0}  e^{-i \phi \hat{n}_1} \hat{\Pi}_1 \ket{\psi}_{1},
\end{equation}
where $\hat{n}_k$  and $\hat{\Pi}_k = e^{i \pi \hat{n}_k}$ denote the photon-number and photon-parity operators for mode $k$, respectively. Owing to the coherent nature of the atom controlling the QM, the optical response of the medium can be modified such that an excited atom $\ket{e}$ leads to complete reflection, whereas its ground state $\ket{g}$ leads to transmission \cite{uria_hermann_solano_delgado_2026,Bekenstein2020}. In this case, the action of a QM is described by
\begin{equation}
\hat{U}_{QM}=\hat{\pi}_g\otimes\mathbf{1}\otimes\mathbf{1} + \hat{\pi}_e\otimes \hat{U}_{M},
\label{eq:UQM}
\end{equation}
where $\hat{\pi}_k = \ketbra{k}{k}$ denotes the projector onto the atomic basis $\{\ket{g},\ket{e}\}$. Here, the phase $\phi$ acquires physical relevance because it governs the superposition between the transmitted and reflected fields.

\begin{figure}[t]
    \centering
    \hspace{-0.1cm}\includegraphics[width=\columnwidth]{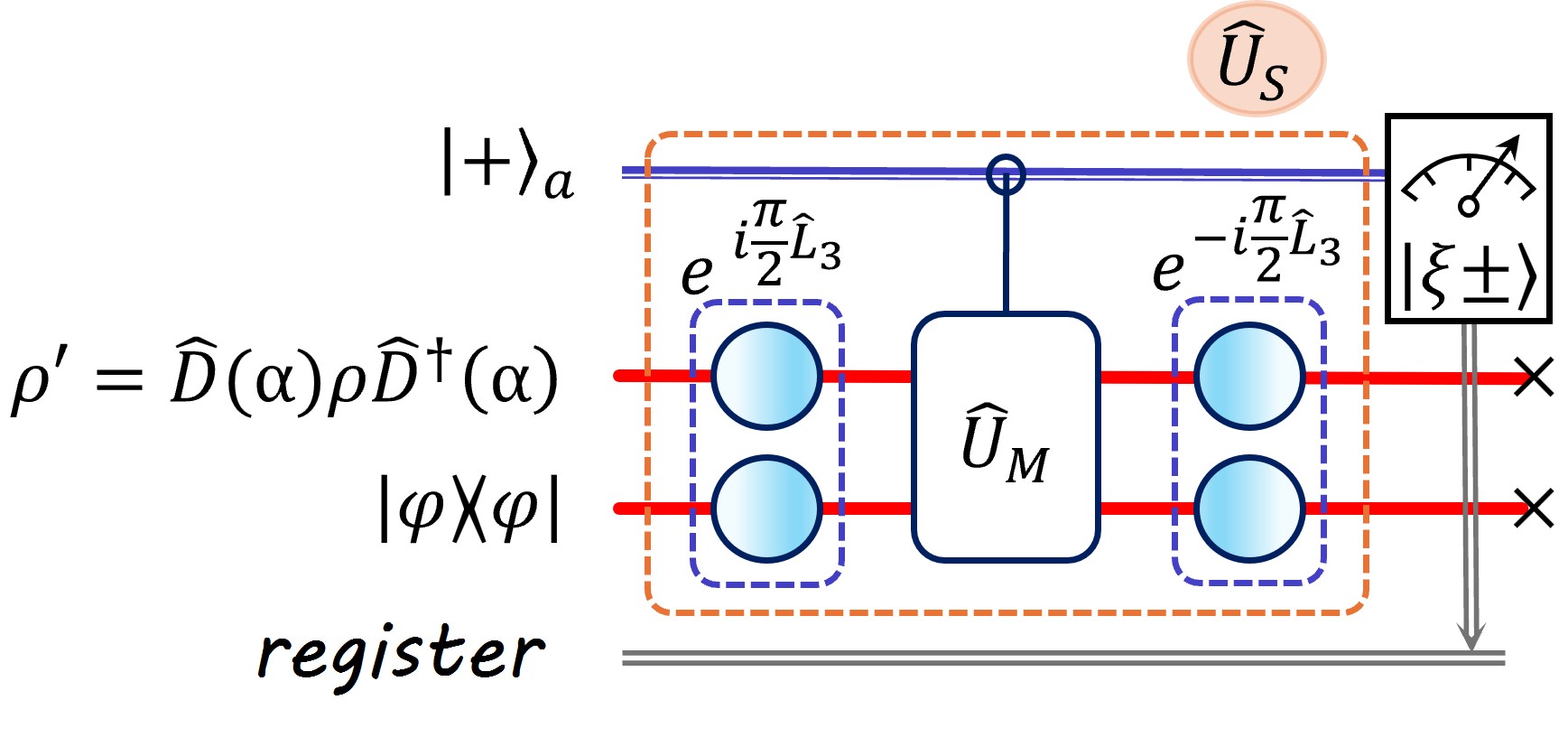}

    \caption{Quantum circuit implementing the unitary operation defined in Eq.~\eqref{eq:UQM}. The system is initialized as an unknown input state $\rho$, coherently displaced by $\alpha$, and a pure probe state $\ket{\varphi}$ in the photonic modes, while the control atom is in a superposition state $\ket{+}$. Finally, the control atom is measured. a measurement is taken onto the atom.}
    \label{fig:qcircuit}
\end{figure}

{\it Measurement of the quantum states of light.---} To reconstruct the quantum state of light, we consider a single QM supplemented by controllable path-dependent phase shifts. Specifically, we apply a phase shift generated by $\hat{L}_3 = (\hat{a}_0^\dagger \hat{a}_0 - \hat{a}_1^\dagger \hat{a}_1)/2$ \cite{Campos1989}, as illustrated in Fig.~\ref{fig:qcircuit}. 

One can engineer a photon-parity operator in the second mode by choosing $\phi=\pi/2$, such that the unitary operator of the setup $\hat{U}_S$ is given by (see Supplementary Material \cite{SM})
\begin{equation} \label{eq:QM}
    \begin{split}
\hat{U}_S\ket{g}_a\ket{\psi}_0\ket{\varphi}_1&=\ket{g}_a\ket{\psi}_0\ket{\varphi}_1\\
\hat{U}_S\ket{e}_a\ket{\psi}_0\ket{\varphi}_1&=| e \rangle_a  \ket{\varphi}_0  \hat{\Pi}_1 \ket{\psi}_1.
    \end{split}
\end{equation}
The platform initializes the control atom in $\ket{+}_a = (\ket{g}_a + \ket{e}_a)/\sqrt{2}$, coherently displaces an input state $\rho$ by $\alpha$, obtaining $\rho' = \hat{D}(\alpha)\rho \hat{D}^\dagger(\alpha)$, and injects a pure state $\ket{\varphi}$ as the probe. The system entangles the control atom and photonic modes via Eq.~\eqref{eq:QM}. Then, measuring the control atom in the convenient superposition states $\ket{\xi_\pm}_a = (\ket{g}_a \pm e^{i\delta}\ket{e}_a)/\sqrt{2}$ maps the interference between the incident fields onto the probabilities (see Supplementary Material \cite{SM})
\begin{equation}
    p_\pm(\delta)=\frac{1}{2}(1\pm \text{Re}[e^{i\delta} \bra{\varphi}\hat{\Pi}\,\hat{D}(\alpha)\rho\hat{D}^{\dagger}(\alpha)\ket{\varphi}]).
    \label{eq:main}
\end{equation}
Here, the photonic modes are traced out, and $\hat{\Pi}$ acts as the photon parity operator. Acknowledging the unusual expression, we note that it relates atomic coherences to the complex value overlap of the photonic states, that can be expressed as
\begin{equation}
\bra{\varphi}\hat{\Pi}\, \hat{D}(\alpha)\rho \hat{D}^\dagger(\alpha)\ket{\varphi}=\left<\hat{\sigma}_x\right>-i\left<\hat{\sigma}_y\right>,
\label{eq: nh}
\end{equation}
which is determined exclusively from measurements of the control atom. Phase $\delta$ provides access to the {\it real} and {\it imaginary} components of this quantity. As we shall see, this expression contains all the information about the state $\rho$.

{\it  $P$ and $Q$ quasiprobability measurements.---} Now, we outline protocols that provide direct access to phase-space representations of $\rho$ via suitable choices of the displacement parameter $\alpha$ and probe states. The Husimi Q function \cite{Agarwal2012}, one of the most widely used quasiprobability distributions due to its favorable analytical and experimental properties \cite{Kirchmair_2013,ahmed_sanchez_munoz_nori_kockum_2021,Leonhardt_1993}, can be measured by choosing the probe state $\ket{\varphi}=\ket{0}$ and the coherent displacement $\alpha=-\gamma$, yielding
\begin{equation}
\bra{\gamma}\rho\ket{\gamma}=p_+(0)-p_-(0),
\end{equation}
which is the Q function, multiplied by $\pi/2$, at the phase space point $\gamma$. This protocol is analogous to the well-known {\it swap} test \cite{Buhrman_2001,Garcia_2013,Volkoff_2022}.

Similarly, the Glauber–Sudarshan $P$ function is the Fourier transform of the matrix element $\bra{-\gamma}\rho\ket{\gamma}$ \cite{Mehta_1967}.  Its negativities or singular features constitute clear signatures of nonclassicality, while its robustness against detection losses \cite{Semenov_2006} makes it a useful witness of quantumness. Simultaneously, its potentially singular character makes experimental reconstruction notoriously challenging \cite{Richter_2001,Kiesel_2008}. Within our protocol, setting the displacement $\alpha=0$ and the probe state $\ket{\varphi}=\ket{\gamma}$ yields the kernel
\begin{equation}
    \bra{-\gamma}\rho\ket{\gamma}=\left<\hat{\sigma}_x\right>-i\left<\hat{\sigma}_y\right>.
\end{equation}
In contrast to the Husimi Q-function case, the parity operator here plays a crucial role, since it provides the $\pi$-rotation of $\gamma$ in $\bra{-\gamma}$.

{\it Wavefunction measurements.---} With minor modifications, we can define a protocol for the direct measurement of the wavefunction of an unknown input pure state $\ket{\psi}$. The protocol starts by choosing a coherent probe state $\ket{\beta/2}$ and coherently displacing the input state by $\beta/2-\gamma$, with $\gamma$ phase-aligned with respect to $\beta$, to simplify the implementation. 

\begin{figure}[t]
    \centering
    \includegraphics[width=0.9\linewidth]{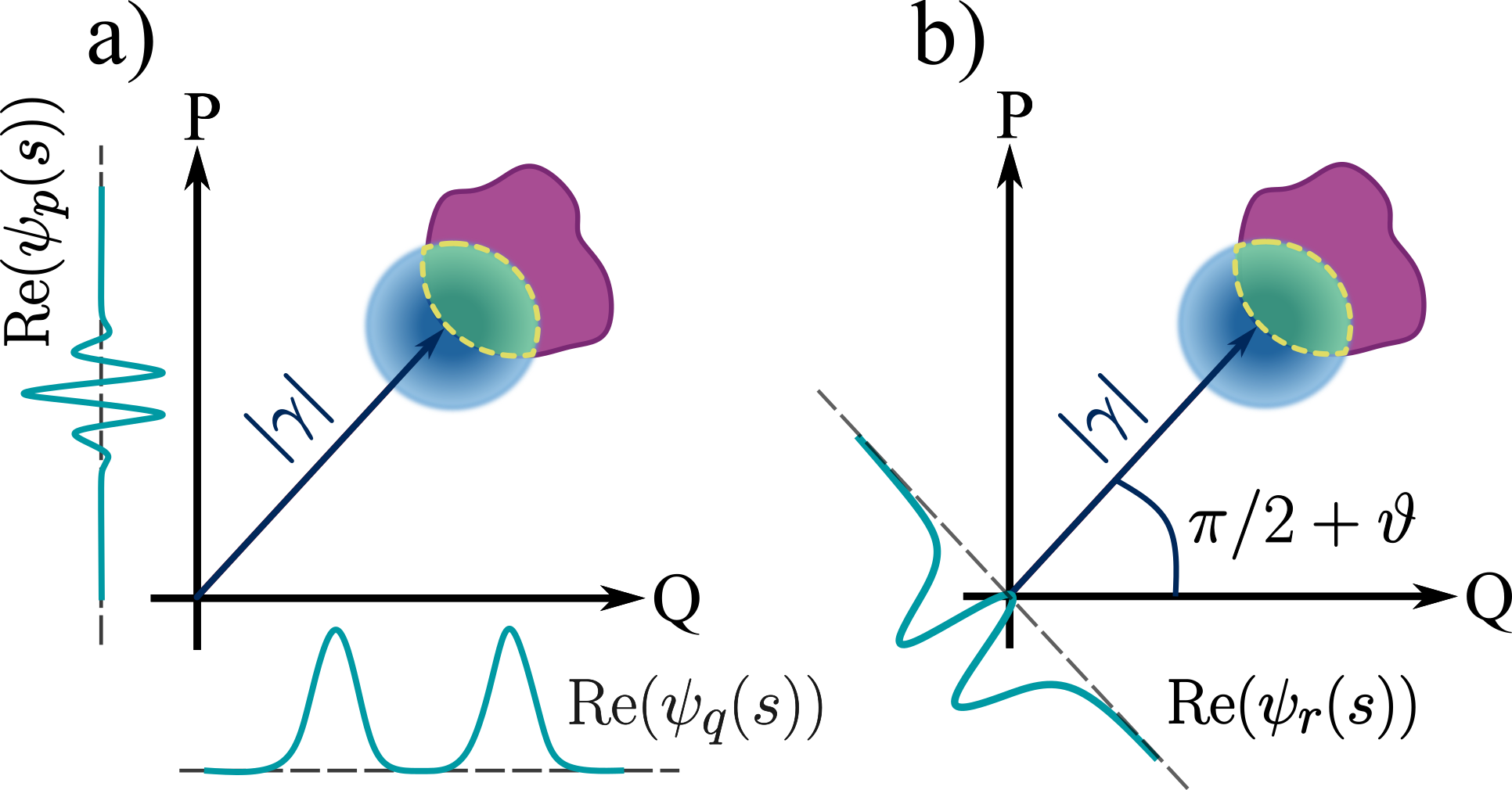}
    \caption{(a) Real part of the projection of the probe state (coherent state in blue) and unknown input state (purple) in the $q$ and $p$ representation. The imaginary part can also be obtained (not shown). (b) Real part of the projection of the probe state (coherent state in blue) and unknown input state (purple) in representation $r$. }
    \label{fig:WF}
\end{figure}

By measuring the control atom according to Eq.~\eqref{eq:main}, one can estimate the value $Z_\beta=\bra{\gamma}\hat{D}(\beta)\ket{\psi}\braket{\psi}{\gamma}$ (see Supplementary Material \cite{SM}). Performing the estimation twice, once for $\beta=0$ and $\beta\neq 0$ one finds
\begin{equation}
\bra{\gamma}\hat{D}(\beta)\ket{\psi}
= \frac{Z_\beta}{\sqrt{Z_0}} e^{-i\phi_0}.
\end{equation}
The wavefunction of $\ket{\psi}$ can be obtained by tuning the parameter $\beta=it \, e^{i\vartheta}/\sqrt{2}$, with $t$ being a real scanning parameter and $\vartheta$ being a conveniently fixed parameter. This quantity can be expressed as the overlap 
\begin{equation}
     \bra{\gamma}\hat{D}(it \, e^{i\vartheta}/\sqrt{2})\ket{\psi} =\int \gamma_r^{*}(s) e^{its} \psi_r(s)ds,
     \label{eq: cross-correlation}
\end{equation}
where  $\psi_r(s)=\braket{r_\vartheta}{\psi}$ and $\gamma_r(s)=\braket{r_\vartheta}{\gamma}$ denote the wavefunctions of the input and probe states, respectively, in the rotated quadrature representation $\ket{r_\vartheta}=e^{i \vartheta \hat{n}}\ket{q}$ which is the eigenstate of $\hat{R}_\vartheta=\hat{Q}\cos{\vartheta}+\hat{P}\sin{\vartheta}$, with the canonical quadrature operators $\hat{Q}=(\hat{a}+\hat{a}^\dagger)/\sqrt{2}$  and $\hat{P}=i(\hat{a}^\dagger-\hat{a})/\sqrt{2}$. 

Equation~(\ref{eq: cross-correlation}) is the inverse Fourier transform \cite{SM,Wolf_1979,Namias_1980} of the product of wavefunctions with respect to variable $t$. Upon applying a Fourier transform to the measurement in $t$, we obtain (see Supplementary Material \cite{SM})
\begin{equation}
F[\bra{\gamma}\hat{D}(i t e^{i\vartheta}/\sqrt{2})\ket{\psi}](\omega)=\sqrt{2\pi}\gamma_r^{*}(\omega)\psi_r(\omega),
\label{eq:wf}
\end{equation}
which directly provides $\psi_r(\omega)$, as $\gamma_r(\omega)$ is known. 

By scanning $\beta$ and applying Eq.~(\ref{eq:wf}), the wavefunction is reconstructed up to the global phase $\phi_0$. A schematic representation of the protocol is shown in Fig.~\ref{fig:WF}, which enables full state characterization directly at the wavefunction level.

{\it Wigner function measurement.---} By adding a second QM, we propose a protocol that enables the direct measurement of the Wigner function for an arbitrary mixed input state $\rho$.

\begin{figure}[tb]
    \centering
    \hspace{-0.75cm}\includegraphics[width=0.85\linewidth]{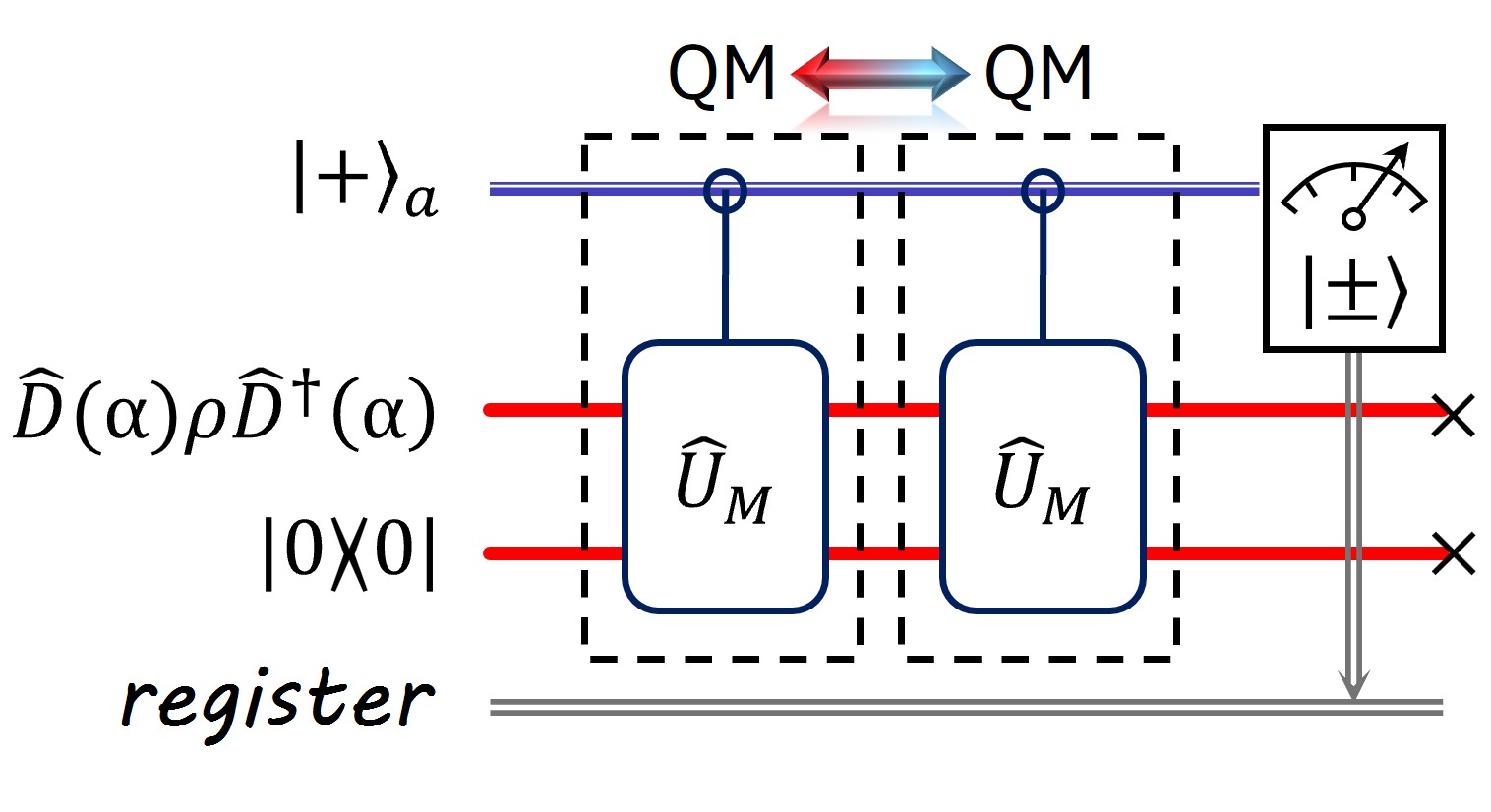}

    \caption{Direct measurement of the Wigner function of the unknown input state $\rho$ at phase space point $\alpha$, represented as a quantum circuit. The setup incorporates a second quantum mirror, which may equivalently be realized by re-injecting the modes into the same QM.}
    \label{fig:qcircuit2}
\end{figure}

The protocol starts by coherently displacing the input state by a known quantity $\alpha$, which defines  a given point in the phase space, while the probe field remains in the vacuum state. As before, the control atom is initialized in the state $\ket{+}_a$, as shown in Fig.~\ref{fig:qcircuit2}. Both quantum mirrors should be controlled using the same atomic state. This can be achieved with a single atom controlling both mirrors, two entangled atoms each controlling a separate mirror, or a single control atom in a single QM upon reinjecting the same photonic modes back into the system. Finally, a measurement is performed on the state $\ket{\pm}_a$, yielding the probabilities
\begin{equation}
    p_\pm=\frac{1}{2}\left(1\pm \text{Tr}\left[\rho \hat{D}^{\dagger}(\alpha) \hat{\Pi}\hat{D}(\alpha)\right]\right).
\end{equation}
Thus, the Wigner function \cite{Moya1993} is explicitly obtained as
\begin{equation}
    W(\alpha) =\frac{2}{\pi}\Big(p_+(\alpha)-p_-(\alpha)\Big).
\end{equation}
This expression has also been obtained in the reconstruction of a field contained in a cavity \cite{Lutterbach_Davidovich_1997}. In our case, the expression above applies to a propagating field and, unlike other approaches, does not require photon counting \cite{Banaszek_Radzewicz_1999}. 

It is worth noting that the Wigner function of a pure state $\rho=\ketbra{\psi}{\psi}$ can also be measured using only a single QM, provided that a copy of the state is introduced as a probe into the setup in Fig.~\ref{fig:qcircuit}. The probe state must be $\ket{\varphi}=\hat{D}(\alpha)\ket{\psi}$, allowing the measurement of the quantity $\bra{\psi}\hat{D}(\alpha) \hat{\Pi}\hat{D}^\dagger(\alpha)\ket{\psi}$, which is proportional to the Wigner function of $\ket{\psi}$.

{\it Experimental feasibility.---} A quantum mirror has been recently demonstrated \cite{Srakaew2023}, where tunable reflectivity was enabled by strong dipolar Rydberg interactions. This provided validation for the theoretical proposals \cite{Shahmoon2017,Bekenstein2020}.

To the best of our knowledge, there are no studies on the relative phase $\phi$ between the reflected and transmitted output fields. This is the last key ingredient for implementing any of the previously mentioned protocols, as detailed in the Supplementary Material~\cite{SM}. Previous studies have shown that the phase of the transmission and reflection coefficients can be modified by engineering the susceptibility of the medium \cite{Ballantine_Ruostekoski_2021}, controlling the population of the relevant atomic levels \cite{Parmee_2021}, considering different geometries \cite{Ballantine_Wilkowski_Ruostekoski_2022, BenMaimon_Solomons_Davidson_Firstenberg_Shahmoon_2025}, or alternatively, by adjusting the polarization and angle of incidence \cite{sun2025tunabledualbandatomicmirror}, which allows for phase modulation while preserving the magnitude of $t$ and $r$.
These works provide a direct route to the experimental control of $\phi$ by tuning the phase of either coefficient. Moreover, this phase can be measured using homodyne detection \cite{Vaneecloo_2022}. Together, these considerations place our proposal within the reach of current experimental capabilities. 

Notably, the protocols presented here are not exclusive to atomically thin metamaterials but can also be realized in nanofibers \cite{Sinha2025, ChangKimble_2012}, conducting nanowires \cite{ChangLukin2007}, and cavity-atom systems \cite{Hetet2011,ReisererRempe_2013,SuZhenXiaoTao_2006}.

{\it Analysis and discussion.---} QMs offer several significant advantages for CV-QST applications. The interference between the input photonic modes is mapped onto the coherences of the atom, which notably enables the measurement of freely propagating photonic states. Moreover, these protocols are robust against phase mismatches between the probe and input photonic fields, as shown in Fig. \ref{fig:qcircuit}.
In addition, the ability to obtain all the information of a field confirms the ability to directly measure the wavefunction, along with the formerly discussed known quasiprobability functions, but is not limited to those \cite{Banaszek_1996}. 

From a technical standpoint, these protocols offer a major advancement in measuring propagating states, comparable to weak measurement protocols \cite{Lundeen2011,ritchie_story_hulet_1991}, 
allowing the characterization of quantum states through the measurement of the control atom. A thorough robustness analysis of the presented protocols is beyond the scope of this study, as it would require details of platform-specific experimental parameters and errors.
Furthermore, this advancement shows promise for characterizing intense/macroscopic quantum fields, which are known for their rapidly scalable complexity in QST protocols, thus opening a unique window for metrology at the Heisenberg limit and for robust quantum information processing.

{\it Conclusions.---} We have developed versatile protocols that leverage quantum mirrors for the direct measurement of quantum states of light, enabling access to a broad range of quasiprobability distributions, as well as the wavefunction itself. By mapping photonic state information onto measurements of a control atom, these methods provide an experimentally feasible approach to continuous-variable quantum state tomography, overcoming the key limitations of traditional techniques, such as photon counting and homodyne tomography. The platform’s flexibility and robustness position it as a promising tool for characterizing complex, high-energy, and non-Gaussian photonic states, with potential applicability extending beyond photonic systems to other quantum technologies, such as superconducting qubits. Thus, this study opens new pathways for efficient benchmarking, verification, and metrology in quantum information science.

\begin{acknowledgments}

{\it Acknowledgments.---} The authors thank K. Sinha for helpful discussions. A.D., C.H., and A.M. acknowledge financial support from the Millennium Science Initiative Program ICN17$_-$012. M. U. was supported by ANID Doctoral Fellowship Grant No. 21232285. C.H. and P.S. acknowledge financial support from FONDECYT REGULAR Grants No. 1230897 and No. 1240204. 

\end{acknowledgments}

\bibliographystyle{apsrev4-2}
\bibliography{references}

\clearpage
\onecolumngrid

\setcounter{page}{1}

\setcounter{section}{0}
\setcounter{equation}{0}
\setcounter{figure}{0}
\renewcommand{\theequation}{S\arabic{equation}}
\renewcommand{\thefigure}{S\arabic{figure}}

\begin{center}
    \textbf{\large SUPPLEMENTARY MATERIAL: \\
    Continuous-Variable Quantum State Tomography Enabled by Quantum Mirrors}
\end{center}

\section{Mode-Independent Beam Splitter Formalism}

Consider a device in which the transmission and reflection coefficients are identical for both input ports (or equivalently, for both propagation directions). The relation between the input and output field modes can be expressed in terms of the symmetric transfer matrix
\begin{equation}
    M  = \begin{pmatrix}
        t & r \\
        r & t
    \end{pmatrix}, 
    \label{eq: matrizsim}
\end{equation}
where $t$ and $r$ denote the transmission and reflection coefficients, respectively. For an ideal (lossless) beam splitter, the input and output modes obey the bosonic commutation relations. Consequently, the matrix $M$ is unitary, leading to the following conservation relations:
\begin{eqnarray} 
|t|^2 + |r|^2 = 1 \label{eq: conservacion1}  \\
 r^*t +  t^*r = 0.\label{eq: conservacion2}    
\end{eqnarray}
In general, these coefficients are complex numbers and can be written as $t=|t|e^{i \phi_{t}}$ and $r=|r|e^{i \phi_{r}}$. Using Eq.~\eqref{eq: conservacion1}  and Eq.~\eqref{eq: conservacion2}, the relation between $r$ and $t$ is often written in the simplified form $t-1 = r$. However, this expression neglects the presence of a \textbf{global phase}, which is commonly considered physically irrelevant but plays a significant role in the present work. With this, the more general relation can be written as
\begin{equation}
   t = r + e^{i \phi}. 
   \label{eq: tr}
\end{equation}
A \textbf{phase relation} for $r$ and $t$ can be derived from Eq.~\eqref{eq: conservacion2}; solving for the phase component
\begin{equation}
        \phi_r-\phi_t = \pm \pi/2.
\end{equation}
A convenient \textbf{parameterization} for $r$ and $t$ can be obtained  from $t = e^{i \phi} + |r| e^{{i\phi_r}}$. Starting from Eq.~\eqref{eq: conservacion1}, one arrives to
\begin{equation*}
        |r| (\cos(\phi-\phi_r) + |r|) = 0,
\end{equation*}
and the solutions for this equation are simply
\begin{equation}
    |r| = 0 \quad \text{and} \quad |r| = - \cos(\phi-\phi_r), 
\end{equation}
where $(\phi_r - \phi )\in [\pi/2, \pi]$ to ensure $|r|\ge0$. Substituting the nontrivial solution into the expression for $t$, we obtain
\begin{equation}
    t = -i e^{i \phi_r} \sin{(\phi_r - \phi)}.
\end{equation}
Using these parameterized coefficients and introducing the change of variables $ \phi_r - \phi = \theta + \frac{\pi}{2}$, $\theta  \in  [0, \pi/2]$. The transfer matrix $M$ can be written as
\begin{equation}
    M = e^{i (\phi+\theta)} 
    \underbrace{
    \begin{pmatrix}
        \cos\theta & i \sin\theta \\
        i \sin\theta & \cos\theta
    \end{pmatrix}}_{{\color{blue}\in \ \text{SU}(2)}}. 
    \label{eq: symmetriccase}
\end{equation}
Since matrix $M$, without the exponential, belongs to the SU(2) group, we adopt the angular-momentum formalism for beam splitter devices, as introduced in Ref.~\cite{Campos1989}. Defining the operators
\begin{equation}
\hat{L}_1 = \frac{1}{2} \left( \hat{a}_0^{\dagger}\hat{a}_1 + \hat{a}_1^{\dagger}\hat{a}_0 \right), \quad
\hat{L}_2 = \frac{1}{2i} \left( \hat{a}_0^{\dagger}\hat{a}_1 - \hat{a}_1^{\dagger}\hat{a}_0 \right), \quad
\hat{L}_3 = \frac{1}{2} \left( \hat{a}_0^{\dagger}\hat{a}_0 - \hat{a}_1^{\dagger}\hat{a}_1 \right),
\label{eq: angular1}
\end{equation}
which leads to the unitary evolution corresponding to the lossless beam splitter associated with Eq.~\eqref{eq: symmetriccase} as
\begin{equation}
    \hat{U}(\theta, \phi)= e^{-i (\theta + \phi)\hat{n} } e^{i \frac{\pi}{2} \hat{L}_3} e^{-2i  \theta \hat{L}_2} e^{-i \frac{\pi}{2} \hat{L}_3} = e^{-i (\theta + \phi)\hat{n} }e^{-2i\theta \hat{L}_1} = e^{-i\left\lbrace (\theta+\phi)\hat{n}+2\theta\hat{L}_1\right\rbrace}, 
    \label{eq:UQMs}
\end{equation}
where $\hat{n}=\hat{a}_0^{\dagger}\hat{a}_0 + \hat{a}_1^{\dagger}\hat{a}_1 $ is the total photon number operator. The corresponding transformation of the field operators is  given by
\begin{eqnarray}
    \hat{a}_0 &\to  \hat{U}(\theta, \phi) \,\hat{a}_0 \,\hat{U}^{\dagger}(\theta, \phi)=e^{i(\theta+\phi)}(\hat{a}_0 \cos\theta+i\hat{a}_1\sin\theta),\\
    \hat{a}_1 &\to  \hat{U} (\theta, \phi) \, \hat{a}_1 \, \hat{U}^{\dagger}(\theta, \phi) =e^{i(\theta+\phi)}(i\hat{a}_0 \sin\theta+\hat{a}_1\cos\theta).
\end{eqnarray}

A particularly relevant case is obtained for $\theta=\pi/2$, which corresponds to a perfect mirror. Defining the mirror operator $\hat{U}_M=\hat{U}(\theta=\pi/2,\phi)$. Its action on an arbitrary two-mode state is
\begin{equation}
   \begin{split}
       \hat{U}_M \ket{\psi}_0\ket{\varphi}_1   &=  \hat{U}_M  \left(\sum_n c_n \frac{\hat{a}_0^{\dagger n}}{\sqrt{n!}}\sum_m d_m \frac{\hat{a}_1^{\dagger m }}{\sqrt{m!}} \right)  \ket{0}\\
       &=\sum_n c_n \frac{(\hat{U}_M \hat{a}_0^{\dagger} \hat{U}_M^{\dagger})^n}{\sqrt{n!}}\sum_m d_m \frac{(\hat{U}_M \hat{a}_1^{\dagger} \hat{U}_M^{\dagger})^m}{\sqrt{m!}} \hat{U}_M \ket{0}
       \\
        &=  \sum_n c_n \frac{(-1)^n(e^{-i \phi n}) (\hat{a}_{1'}^{\dagger})^n}{\sqrt{n!}} \sum_m d_m \frac{(-1)^m(e^{-i \phi m}) (\hat{a}_{0'}^{\dagger})^m}{\sqrt{m!}} \ket{0} \\
        &=e^{-i \phi \hat{n}_{0'}}\hat{\Pi}_{0'} \ket{\varphi}_{0'} e^{-i \phi \hat{n}_{1'}} \hat{\Pi}_{1'} \ket{\psi}_{1'}
   \end{split} 
   \label{eq:QMket}
\end{equation}
where $\hat{\Pi}_k = e^{i \pi \hat{n}_k}$ is the photon-parity operator associated with mode $k$ and $\ket{n}$ is the Fock basis. Hence, each reflected field acquires the same phase shift $\phi$ and a parity operation. At this stage, the factors $e^{-i\phi\hat{n}_k}$ contribute only global phases to the individual modes and are therefore physically irrelevant in the absence of a relative phase.

\section{Parity operator manipulation}

We define the unitary operation of the \textit{quantum mirror} (QM) as a perfect mirror coherently controlled by a two-level system
\begin{equation}    \hat{U}_{QM}=\hat{\pi}_g\otimes\mathbf{1}\otimes\mathbf{1} + \hat{\pi}_e\otimes \hat{U}_{M},
    \label{eq:UQM2}
\end{equation}
where $\hat{\pi}_k = \ketbra{k}{k}$ denotes the projector onto the atomic basis $\{\ket{g},\ket{e}\}$ and the operator $\hat{U}_{M}$ is defined as in Eq.~\eqref{eq:QMket}.
In contrast to the previous situation, the phase factors associated with $\hat{U}_M$ are no longer global, as they appear as relative phases imprinted on the reflected-field modes conditioned on the atomic states $\ket{g}$ and $\ket{e}$. Consequently, these phases become physically observable through interference in the atomic degree of freedom. 

\begin{figure}[ht]
    \centering
    \hspace{-3cm}\includegraphics[width=0.5\linewidth]{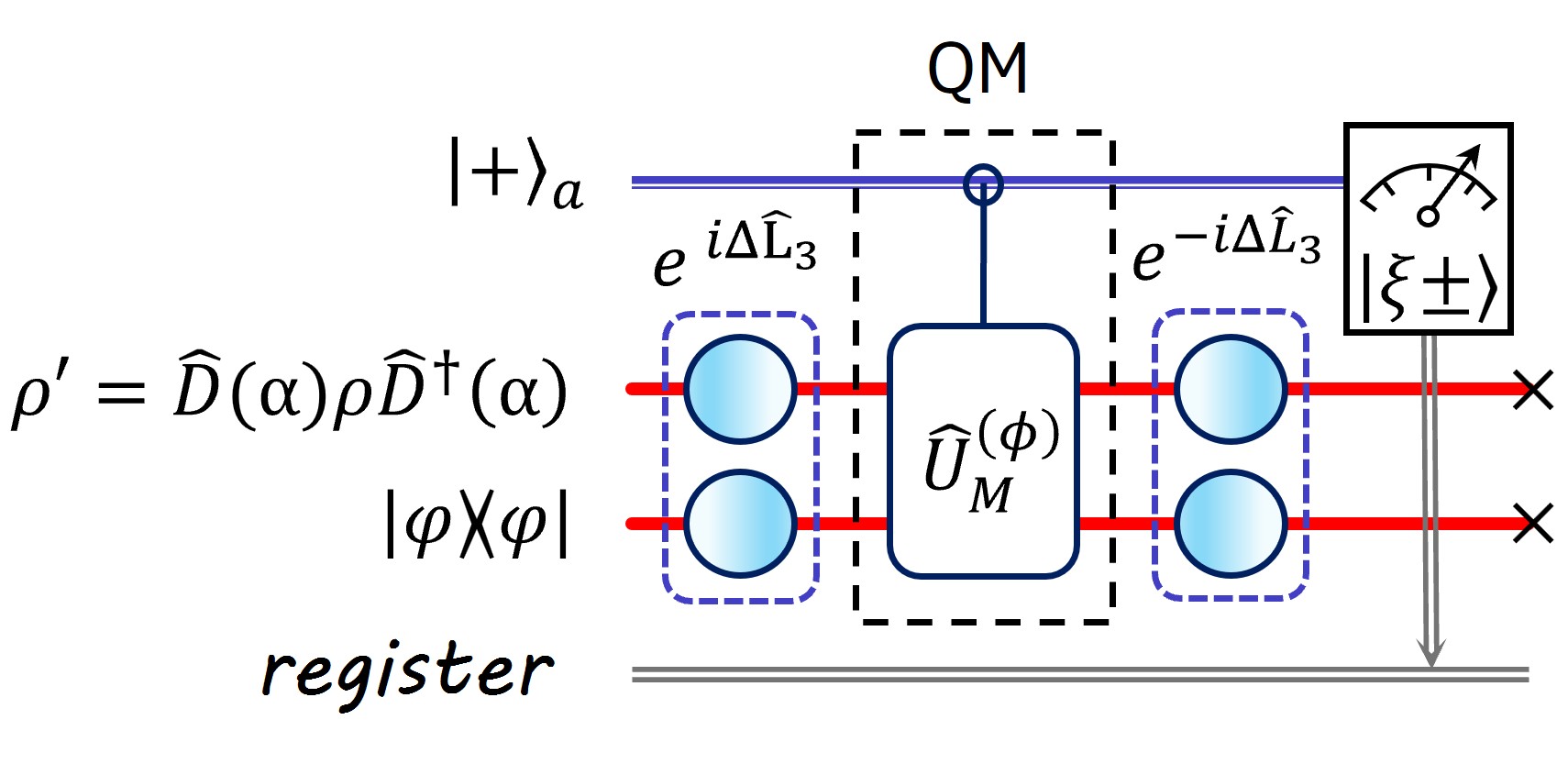}
    \caption{Quantum circuit implementing the unitary operator defined in Eq.~\eqref{eq:system} The phase symmetrical response of the quantum mirror (QM) is modified by introducing relative path-dependent phases $\Delta$ before and after the QM, enabling the existence of a photon-parity operator acting on only one of the two modes. The mode in which this parity operator is applied depends on the choice of $\Delta$ and $\phi$. The operator $\hat{U}_M^{(\phi)}$ becomes $\hat{U}_S$ once the phase $\phi=\pi/2$ is set, as shown in Fig. \ref{fig:qcircuit}. }
    \label{fig:phases}
\end{figure}

Additional control over these phases can be achieved by introducing mode-dependent phase shifts $\{\varphi_0,\varphi_1,\varphi_2,\varphi_3\}$ before and after the QM interaction. These phases can be physically implemented using optical path-length differences or external phase shifters that act independently on each mode. The corresponding unitary operations are
\begin{equation}
e^{i(\varphi_0\hat{n}_0+\varphi_1\hat{n}_1)}
\qquad\text{and}\qquad
e^{-i(\varphi_2\hat{n}_0+\varphi_3\hat{n}_1)},
\end{equation}
respectively. The resulting system operator is therefore
\begin{equation}
    \begin{split}
        \hat{U}_S  =&\, e^{-i(\varphi_2 \hat{n}_0 + \varphi_3 \hat{n}_1)} \,\hat{U}_{QM}\, e^{i(\varphi_0 \hat{n}_0 + \varphi_1 \hat{n}_1)},\\
        =& \, \hat{\pi}_g \otimes e^{i (\varphi_{in} - \varphi_{out}) \hat{n}/2} e^{i (\Delta  - \Delta') \hat{L}_3} \\
  &+ \hat{\pi}_e \otimes  e^{ i (\varphi_{in} - \varphi_{out}) \hat{n}/2} e^{- i \Delta' \hat{L}_3} \hat{U}_{M}\,  e^{ i \Delta \hat{L}_3},
\label{eq:aaa1}
    \end{split}
\end{equation}
where we have introduced the phase sums and differences $\varphi_{in} = \varphi_0 + \varphi_1 $, $\varphi_{out} = \varphi_2 + \varphi_3 $, $\Delta  = \varphi_0 - \varphi_1$ and $\Delta' = \varphi_2  -\varphi_3$. Factorizing the common terms yields
\begin{align}
  \hat{U}_S  =  \mathbf{1}\otimes e^{i (\varphi_{in} - \varphi_{out}) \hat{n}/2} e^{i (\Delta - \Delta') \hat{L}_3}  \left( \hat{\pi}_g \otimes \mathbf{1}\otimes\mathbf{1} 
  + \hat{\pi}_e \otimes   e^{- i \Delta \hat{L}_3}\, \hat{U}_{M}\,  e^{ i \Delta \hat{L}_3} \right). 
\label{eq:aaa2}
\end{align}
Without loss of generality, we set $\varphi_0=\varphi_2$ and $\varphi_1=\varphi_3$, because the factorized term corresponds to a global operation on the field modes and has no physical consequence when only the atomic subsystem is measured (see Fig.~\ref{fig:phases}). Then the simplified unitary system operator is given by
\begin{equation}
    \hat{U}_S  =  \hat{\pi}_g \otimes  \mathbf{1} \otimes \mathbf{1} 
  + \hat{\pi}_e \otimes  e^{- i \Delta \hat{L}_3}\, \hat{U}_{M} \, e^{ i \Delta \hat{L}_3}.
  \label{eq:system}
\end{equation}
Finally, the action of $\hat{U}_S$ on two incoming field modes prepared in arbitrary states $\ket{\psi}$ and $\ket{\varphi}$ and on the atomic state is given by
\begin{equation}
    \begin{split}
\hat{U}_S\ket{g}\ket{\psi}\ket{\varphi}&=\ket{g}\ket{\psi}\ket{\varphi}\\
\hat{U}_S\ket{e}\ket{\psi}\ket{\varphi}&=| e \rangle  e^{-i (\phi+\Delta) \hat{n}_0} \hat{\Pi}_0 \ket{\varphi} \otimes e^{-i (\phi-\Delta) \hat{n}_1} \hat{\Pi}_1 \ket{\psi}.
    \end{split}
\end{equation}
Then, by appropriately tuning the phases $\phi$ and $\Delta$ one can choose the mode in which a parity operator is applied.

The platform consists of initializing the control atom in $\ket{+}_a = (\ket{g}_a + \ket{e}_a)/\sqrt{2}$,
an input state $\rho$ displaced by $\alpha$, obtaining $\rho' = \hat{D}(\alpha)\rho \hat{D}^\dagger(\alpha)$, and a pure state $\ket{\varphi}$ as the probe. As shown in Fig.~\ref{fig:phases}, the initial state of the system can be written as
\begin{equation} \label{eq:systeminit}
    \rho_{sys}^{(i)} = \ketbra{+}{+} \otimes \rho' \otimes \ketbra{\varphi}{\varphi} = \left(\frac{\hat{\pi}_{g}+\hat{\pi}_{ge}+\hat{\pi}_{eg}+\hat{\pi}_{e}}{2}\right) \otimes \left( \sum_{i} p_{i}' \ketbra{\psi_{i}'}{\psi_{i}'} \right) \otimes \ketbra{\varphi}{\varphi}.
\end{equation}
The action of the operator $\hat{U}_S$, defined in Eq.~\eqref{eq:systeminit}, on the initial state is given by 
\begin{equation}
    \begin{split}
        &\hat{U}_{S} \,\rho_{sys}^{(i)}\, \hat{U}_{S}^{\dagger} =\frac{1}{4} (
         \pi_{gg} \otimes
           \rho' 
         \otimes
           \ket{\varphi}\bra{\varphi}  +
         \pi_{ge} \otimes 
          \sum_{i} p_{i}' \ket{\psi_{i}'} \bra{\varphi}  \hat{U}^{\dagger}_{\pi +\Delta - \phi }
        \otimes
          \ket{\varphi} \bra{\psi_{i}'}  
        \hat{U}^{\dagger}_{\pi - \Delta - \phi }  \\+
        &\pi_{eg} \otimes 
         \sum_{i}p_{i}' \,  \hat{U}_{\pi -\Delta - \phi }
         \ket{\varphi}\bra{\psi_{i}'}  
        \otimes 
         \hat{U}_{\pi + \Delta - \phi } \ket{\psi_{i}'}\bra{\varphi} + 
        \pi_{ee} \otimes
         \hat{U}_{ \pi -\Delta - \phi} \ket{\varphi}\bra{\varphi} \hat{U}^{\dagger}_{ \pi - \Delta - \phi}
        \otimes
         \hat{U}_{ \pi + \Delta - \phi} 
         \rho'  \hat{U}^{\dagger}_{ \pi + \Delta - \phi}),
    \end{split} \label{eq:operationwithphases}
\end{equation}
where $\hat{U}_\alpha=e^{i\alpha\hat{n}}$ implements a phase shift $\alpha$ in the corresponding mode.
We then apply a partial trace over the photonic modes
\begin{equation} 
    \begin{split}
         \rho_{at}^{(f)}&=\text{Tr}_{0,1}\{\hat{U}_{S} \rho
     _{sys} \hat{U}_{S}^{\dagger}\} \\
     &= \frac{1}{4}\Big(\hat{\pi}_{gg} 
     + 
     \hat{\pi}_{eg}  \sum_i p_i'\bra{\psi_i'} \hat{\Pi}_0 e^{-i(\phi + \Delta)n_0}    \ket{\varphi} \bra{\varphi} \hat{\Pi}_1 e^{-i(\phi- \Delta)n_1} \ket{\psi_i'} 
     \\
     &\qquad + \hat{\pi}_{ge} \sum_i p_i' \bra{\varphi} \hat{\Pi}_0^{\dagger} e^{i(\Delta + \phi)n_0}    \ket{\psi_{i}'} \bra{\psi_i'} \hat{\Pi}_1^{\dagger} e^{i(\phi- \Delta)n_1}  \ket{\varphi} + \hat{\pi}_{ee}\Big)\\
     &=\frac{1}{4}\Big(\hat{\pi}_{gg} 
     + 
     \hat{\pi}_{eg}  \bra{\varphi} \hat{\Pi}\,  e^{-i(  \phi-\Delta)\hat{n}}  \hat{D}(\alpha) \rho \hat{D}^{\dagger}(\alpha) \hat{\Pi}\, e^{-i( \phi + \Delta)\hat{n}}  \ket{\varphi} 
     \\
     &\quad\qquad + \hat{\pi}_{ge} \bra{\varphi} \hat{\Pi}^{\dagger}  e^{i(  \Delta + \phi)\hat{n}}  \hat{D}(\alpha) \rho \hat{D}^{\dagger}(\alpha) \hat{\Pi}^{\dagger} e^{i( \phi - \Delta)\hat{n}}  \ket{\varphi}+ \hat{\pi}_{ee}\Big),
    \end{split}
\end{equation}
which allows us to recognize the interference terms on the final state of the control atom. Measuring them on the superposition basis $\ket{\xi_\pm}_a = (\ket{g}_a \pm e^{i\delta}\ket{e}_a)/\sqrt{2}$, we can write the probabilities as 
\begin{equation}   
    \boxed{ p_{\pm} (\delta)= \frac{1}{2} \left(1 \pm  \text{Re} \left[e^{i \delta}   \bra{\varphi} \hat{\Pi}^{\dagger}  e^{i(  \Delta + \phi)\hat{n}}  \hat{D}(\alpha) \rho \hat{D}^{\dagger}(\alpha) \hat{\Pi}^{\dagger} e^{i( \phi - \Delta)\hat{n}}  \ket{\varphi} \right]\right)}
\end{equation}
In the main manuscript, we have selected the parameters $\phi=\Delta=\pi/2$.

\section{Wavefunction reconstruction}

The electromagnetic field can be described as a collection of decoupled harmonic oscillators, each characterized by canonical operators $position$ $\hat{q}$ and $momentum$ $\hat{p}$ satisfying the commutation relation $[\hat{q},\hat{p}]=i\hbar$. For each mode of frequency $\omega$, the corresponding bosonic annihilation operator is defined as
\begin{equation}
    \hat{a}=\sqrt{\frac{1}{2\hbar\omega}}(\omega\hat{q}+i\hat{p}).
\end{equation}
It is convenient to introduce the dimensionless quadrature operators $\hat{Q}=\hat{q}\sqrt{\omega/\hbar}$ and $\hat{P}=\hat{p}/\sqrt{\hbar\omega}$, which satisfy the normalized commutation relation $[\hat{Q},\hat{P}]=i$. Since $\hat{Q}$ is proportional to $\hat{q}$, they both share the same eigenstates in the continuous spectra $\ket{q}$ such that $\hat{Q}\ket{q}=s\ket{q}$ with $s=q\sqrt{\omega/\hbar}$.

On the other hand, one may generate a rotated quadrature by applying the phase-space rotation operator $e^{i \vartheta \hat{n}}$, acting on the eigenvalue equation for $\hat{Q}$

\begin{equation}
    \left(\frac{\hat{a}e^{-i \vartheta}+\hat{a}^\dagger e^{i \vartheta}}{\sqrt{2}}\right)\ket{r_{\vartheta}}=\underbrace{(\hat{Q}\cos \vartheta +\hat{P} \sin \vartheta)}_{\color{blue}\hat{R}_{\vartheta}}\ket{r_{\vartheta}}=s\ket{r_{\vartheta}},
\label{eq:R}
\end{equation}
where $\ket{r_\vartheta}=e^{i\vartheta \hat{n}}\ket{q}$ is the eigenstate of the rotated quadrature operator $\hat{R}_\vartheta$, which retains the same eigenvalue $s$. In the particular case of $\vartheta=\pi/2$, one finds $\ket{r_{\pi/2}}=e^{i\pi\hat{n}/2}\ket{q}=\ket{p}$ which is the eigenstate of the momentum-like quadrature operator $\hat{P}$.

We now derive the wavefunctions of Fock states in the $q$-representation. Starting from the eigenvalue equation $\hat{a}^\dagger\hat{a}\ket{n}=n\ket{n}$ and projecting onto the continuous basis $\ket{q}$, one obtains the differential equation 
\begin{equation}
 \frac{\partial^2 \psi_q^{(n)}(s)}{\partial s^2}+(2n+1-s^2)\psi_q^{(n)}(s)=0,
\end{equation}
where $\psi_q^{(n)}(s)=\braket{q}{n}$. The normalized solutions in $s$ variable are given by
\begin{equation}
    \psi_q^{(n)}(s)=\frac{1}{(2^n n!)^{1/2}}\sqrt[4]{\frac{1}{\pi}}e^{-s^2/2}\text{H}_n(s),
\end{equation}
where $H_n(s)$ denotes the Hermite polynomial of order $n$.

The wavefunction of a coherent state $\ket{\alpha}$ in the $q$-representation can be obtained from its expansion in the Fock basis and projecting onto $\ket{q}$. Then, using the generating function of Hermite polynomials, one finds
\begin{equation}
\begin{split}
\psi_q^{(\alpha)}(s)=\braket{q}{\alpha}&=e^{-|\alpha|^2/2}\sum_{n=0} \frac{\alpha^n}{\sqrt{n!}}\psi_s^{(n)}(s)=\sqrt[4]{\frac{1}{\pi}}e^{-|\alpha|^2/2}e^{-s^2/2}\sum_{n=0} \frac{(\alpha/\sqrt{2})^n}{n!}\text{H}_n(s)\\
&=\sqrt[4]{\frac{1}{\pi}}e^{-|\alpha|^2/2}e^{-s^2/2}e^{-\alpha^2/2+\sqrt{2}\alpha s}
\end{split}
\label{eq:alphaq}
\end{equation}

Since the $q$-representation and $p$-representation wavefunctions are related by a Fourier transform, we introduce the following definition \cite{Wolf_1979}:
\begin{eqnarray}
    \tilde{f}(k)&=\frac{1}{\sqrt{2\pi}}\int_{-\infty}^{\infty}f(s) \text{ e}^{-isk}ds:=(\hat{F}f)(k) \label{eq:Fourier}\\
    f(s)&=\frac{1}{\sqrt{2\pi}}\int_{-\infty}^{\infty}\tilde{f}(k)\text{ e}^{isk}dk:=(\hat{F}^{-1}f)(s) \label{eq:invFourier}
\end{eqnarray}
where an operator form $\hat{F}$ of this transform is introduced. Applying this operator to the wavefunction of a Fock state in the $q$-representation yields
\begin{equation}
    \underbrace{\psi_p^{(n)}(s)}_{\color{red}\braket{p}{n}}=\underbrace{\hat{F}\psi_q^{(n)}(s)}_{\color{red}\bra{q}e^{-i\frac{\pi}{2} \hat{n}}\ket{n}}=e^{-i\pi n/2}\psi_q^{(n)}(s)=(-i)^n \psi_q^{(n)}(s),
\end{equation}
where the Fourier transform operator can be expressed in terms of the photon number operator as $\hat{F}=e^{-i\frac{\pi}{2} \hat{n}}$. 

Because the $q$-representation wavefunctions belong to the square-integrable space $L^2(\mathbb{R})$, the Fourier transform is a closed operation and maps this space onto itself.

As an example, we determine the wavefunction of a coherent state $\ket{\alpha}$ in the $p$-representation. Although this result can be obtained in several ways, the most straightforward approach employs the operator representation of the Fourier transform
\begin{equation}
    \psi_p^{(\alpha)}(s)=\braket{p}{\alpha}=\bra{q}e^{-i\frac{\pi}{2} \hat{n}}\ket{\alpha}=\braket{q}{-i\alpha}=\sqrt[4]{\frac{1}{\pi}}e^{-|\alpha|^2/2}e^{-s^2/2}e^{\alpha^2/2-i\alpha s\sqrt{2}}.
\end{equation}
This result can be independently verified by directly applying the Fourier transform defined in Eq.~\eqref{eq:Fourier} to the explicit $q$-representation wavefunction given in Eq.~\eqref{eq:alphaq}.

The above reasoning can be extended beyond the $q$ and 
$p$ representations to a continuous family of intermediate quadrature representations obtained by an arbitrary phase-space rotation generated by the number operator. Defining the rotated quadrature eigenstates as $\ket{r_{\vartheta}}=e^{i \vartheta \hat{n}}\ket{q}$, the corresponding wavefunction of the coherent state reads as
\begin{equation}
\begin{split}
        \psi_r^{(\alpha)}(s)&=\braket{r_{\vartheta}}{\alpha}=\bra{q}e^{-i\vartheta \hat{n}}\ket{\alpha}=\braket{q}{\alpha e^{-i\vartheta}}=\sqrt[4]{\frac{1}{\pi}}e^{-|\alpha|^2/2}e^{-s^2/2}e^{-\alpha^2e^{-2i\vartheta}/2+\alpha s e^{-i\vartheta} \sqrt{2}}\\
    &=F_{-\vartheta}[\psi_q^{(\alpha)}(s)],
\end{split}
\end{equation}
where the rotation operation corresponds to a fractional Fourier transform (FrFT) that maps the $q$-representation to an arbitrary quadrature representation. The FrFT is defined as \cite{Namias_1980}
\begin{equation}
    [F_\xi f](k)=\frac{e^{i(\pi-2\xi)/4}}{\sqrt{2\pi \sin\xi}}e^{-ik^2\cot(\xi)/2}\int_{-\infty}^{\infty}e^{-is^2\cot(\xi)/2+isk/\sin\xi} f(s)ds.
\end{equation}
For $\xi=-\pi/2$, this expression reduces to the Fourier transform defined in Eq.~\eqref{eq:Fourier}.

\subsection{Cross-correlation procedure}
Using an appropriate choice of variables, the displacement operator can be rewritten as a function of quadrature operators
\begin{equation}
    \hat{D}\left(\frac{it e^{i\vartheta}}{\sqrt{2}}\right)=e^{it(\hat{Q}\cos\vartheta+\hat{P}\sin\vartheta)}=e^{it\hat{R}_{\vartheta}}.
\end{equation}
Using this representation, the wavefunction of the unknown state $\ket{\psi}$ can be reconstructed through a cross-correlation with a reference coherent state $\ket{\alpha}$. Upon applying an appropriate completeness operation, the inverse Fourier transform of the product wavefunctions naturally arises.  Subsequently, the bare product of the wavefunctions is recovered by applying the Fourier transform. The details of this procedure are described below for two cases.
\\

\noindent\underline{Wavefunction product in $q$-representation ($\vartheta=0$)}
\begin{equation}
\bra{\alpha}\hat{D}\left({it}/{\sqrt{2}}\right)\ket{\psi}=\int \bra{\alpha}e^{it\hat{Q}}\ket{q}\braket{q}{\psi}ds=\int \braket{\alpha}{q}e^{its}\braket{q}{\psi}ds=\int \psi_q^{(\alpha)*}(s) e^{its} \psi_q(s)
\end{equation}
\begin{equation}
F\left[\bra{\alpha}\hat{D}\left({it}/{\sqrt{2}}\right)\ket{\psi}\right](\omega)=\frac{1}{\sqrt{2\pi}}\iint \psi_q^{(\alpha)*}(s)\psi_q(s)e^{its}e^{-i\omega t}ds dt=\frac{1}{\sqrt{2\pi}}\int \psi_q^{(\alpha)*}(s)\psi_q(s)\underbrace{\int e^{it(s-\omega)}dt}_{\color{red}2\pi\delta(s-\omega)} ds
\end{equation}
\begin{equation}  F\left[\bra{\alpha}\hat{D}\left({it}/{\sqrt{2}}\right)\ket{\psi}\right](\omega)= \sqrt{2\pi} \psi_q^{(\alpha)*}(\omega)\psi_q(\omega)
\end{equation}
\\

\noindent\underline{Wavefunction product in $p$-representation ($\vartheta=\pi/2$)}
\begin{equation}
    \bra{\alpha}\hat{D}\left(-{t}/{\sqrt{2}}\right)\ket{\psi}=\int \bra{\alpha}e^{it\hat{P}}\ket{p}\braket{p}{\psi}ds=\int \braket{\alpha}{p}e^{its}\braket{p}{\psi}ds=\int \psi_p^{(\alpha)*}(s) e^{its} \psi_p(s)
\end{equation}

\begin{equation}
  F\left[\bra{\alpha}\hat{D}\left(-{t}/{\sqrt{2}}\right)\ket{\psi}\right](\omega)= \sqrt{2\pi} \psi_p^{(\alpha)*}(\omega)\psi_p(\omega)
\end{equation}
\\

\noindent\underline{Wavefunction product in $r$-representation}: In this particular case, we choose the phase $\vartheta = \delta - \pi/2$ such that the phase of the displacement operator coincides with the phase $\delta$ of the reference coherent state $\ket{\alpha}$, which provides an experimentally convenient implementation due to the shared phase.
\begin{equation}
    \bra{\alpha}\hat{D}\left({t e^{i\delta}}/{\sqrt{2}}\right)\ket{\psi}=\int \bra{\alpha}e^{it\hat{R}_{\vartheta}}\ket{r_{\vartheta}}\braket{r_{\vartheta}}{\psi}ds=\int \psi_r^{(\alpha)*}(s) e^{its} \psi_r(s)
\end{equation}

\begin{equation}
 F\left[\bra{\alpha}\hat{D}\left({t e^{i\delta}}/{\sqrt{2}}\right)\ket{\psi}\right](\omega)= \sqrt{2\pi} \psi_r^{(\alpha)*}(\omega)\psi_r(\omega)
\end{equation}
An alternative choice is $\vartheta = \delta$, for which the coherent state wavefunction is real in that representation; however, this option requires delicate experimental control of the phase in the orthogonal quadrature for displacement.

\subsection{Quantum mirror implementation}

As shown in the main paper, the quantity measurable by quantum mirrors is $\bra{\varphi}e^{i\pi\hat{n}}\hat{D}(\alpha)\rho\hat{D}(\alpha)\ket{\varphi}$. In order to reconstruct the wavefunction, it is necessary to measure the cross-correlation between a coherent state $\ket{\gamma}$ and the unknown pure state $\ket{\psi}$, namely
\begin{equation}
    \bra{\gamma}\hat{D}(\beta)\ket{\psi},
\end{equation}
where $\beta$ contains the variable dependencies, as discussed in the previous subsection. Therefore, the measurement procedure was performed in two stages.

We choose the probe state as the coherent state $\ket{\beta/2}$ and displace the unknown state by $\beta/2-\gamma$. The measurable quantity then becomes
\begin{equation}
\begin{split}
    Z_\beta
    &= \bra{\beta/2} e^{i\pi\hat{n}}
    \hat{D}(\beta/2-\gamma)\ket{\psi}
    \bra{\psi}
    \hat{D}^\dagger(\beta/2-\gamma)
    \ket{\beta/2} \\
    &= \bra{-\beta/2}
    \hat{D}(\beta/2-\gamma)\ket{\psi}
    \braket{\psi}{\gamma} \\
    &= \bra{0}
    \hat{D}(-\gamma)\hat{D}(\beta)\ket{\psi}
    \braket{\psi}{\gamma} \\
    &= \bra{\gamma}
    \hat{D}(\beta)\ket{\psi}
    \braket{\psi}{\gamma}.
\end{split}
\end{equation}
In particular, $Z_0 = |\braket{\psi}{\gamma}|^2$, which implies $\braket{\psi}{\gamma} = \sqrt{Z_0}\,e^{i\phi_0}$. Therefore, the desired overlap can be extracted as
\begin{equation}
    \bra{\gamma}\hat{D}(\beta)\ket{\psi}
    = \frac{Z_\beta}{\sqrt{Z_0}}e^{-i\phi_0}.
\end{equation}
Finally, the wavefunction in the $r$-representation can be reconstructed according to
\begin{equation}
    \sqrt{2\pi}\,
    \psi_r^{(\gamma)*}(\omega)\psi_r(\omega)
    =
    F\!\left[
    \bra{\gamma}\hat{D}(\beta)\ket{\psi}
    \right]
    =
    \frac{e^{-i\phi_0}}{\sqrt{Z_0}}
    F[Z_\beta],
\end{equation}
where $F$ denotes the Fourier transform. If a different quadrature representation is required, the corresponding fractional Fourier transform (FrFT) can be applied to the signal.

\section{Direct measurement of the Wigner function}

\begin{figure}[ht]
\centering
    \hspace{-2cm}\includegraphics[width=0.4\linewidth]{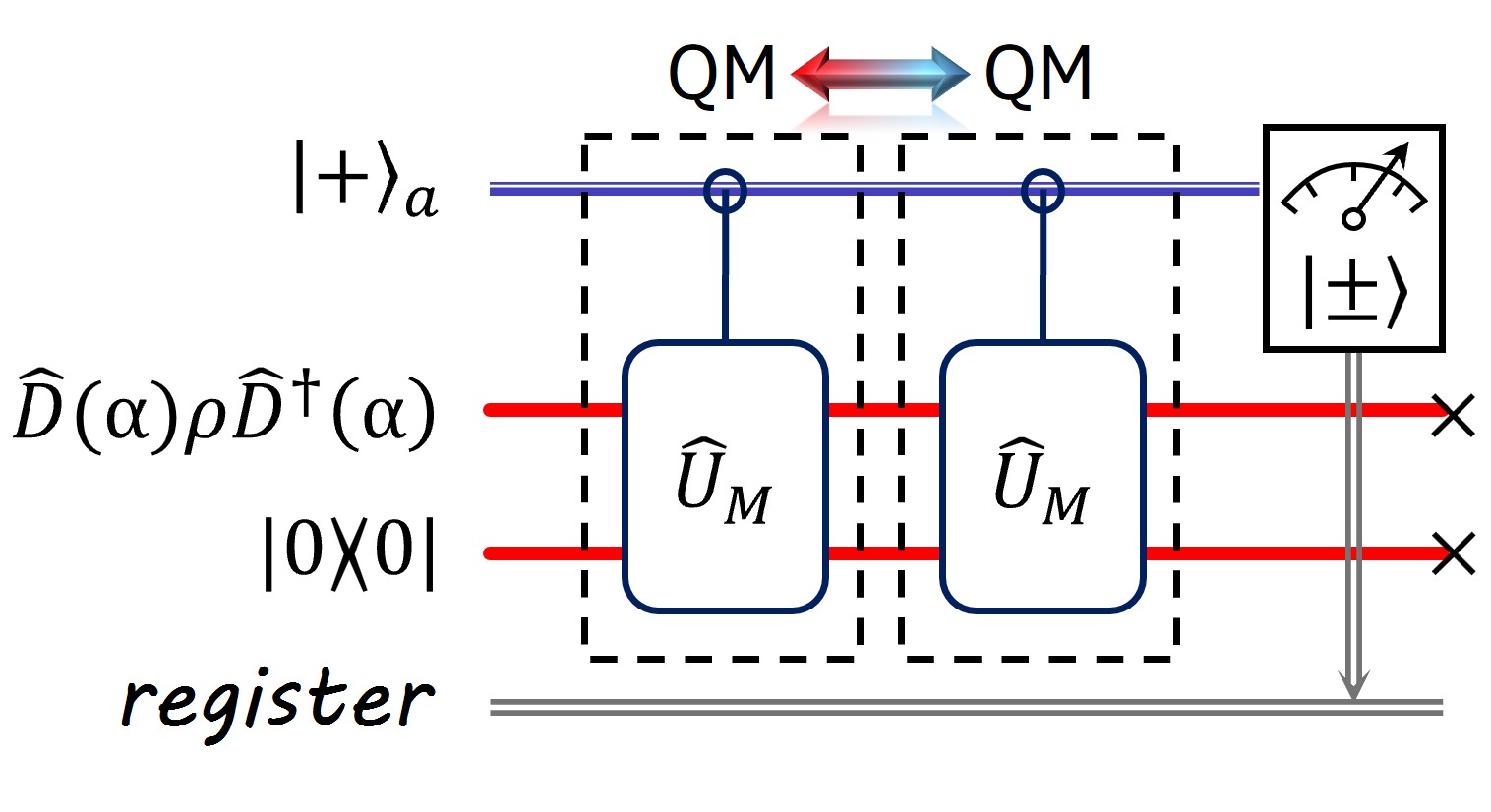}
    \caption{Direct measurement of the Wigner function of an unknown state $\rho$ at phase space point $\alpha$, represented as a quantum circuit. The setup incorporates a second QM, which may equivalently be realized by re-injecting the modes into the same QM.}
    \label{fig:wigner}
\end{figure}

The Wigner function of the state $\rho$ can be measured by applying two QMs, as shown in Fig.~\ref{fig:wigner}. The corresponding unitary operator of this system is given by
\begin{equation}
    \hat{U}_{QM} \cdot \hat{U}_{QM} = \hat{\pi}_g \otimes \mathbf{1} \otimes \mathbf{1} + \hat{\pi}_e \otimes \hat{U}_{M}^2.
\end{equation}
We now evaluate the action of this operator on two incoming field modes prepared in arbitrary states $\ket{\psi}$ and $\ket{\varphi}$, assuming that the atom is initially in the excited state $\ket{e}$:
\begin{equation}
    \begin{split}
        \hat{U}_{QM}  \cdot \hat{U}_{QM} \ket{e} \ket{\psi} \ket{\varphi} &=  \hat{U}_{QM} \cdot\left( \ket{e} \otimes  e^{-i \phi \hat{n}_0 } \hat{\Pi}_0 \ket{\varphi} \otimes  e^{-i \phi \hat{n}_1 } \hat{\Pi}_1 \ket{\psi}
     \right)\\
     &= \ket{e} \otimes e^{ -2i  \phi \hat{n}_0 }  \ket{\psi} \otimes e^{ -2i  \phi \hat{n}_1 } \ket{\varphi}.
    \end{split}
\end{equation}

Thus, the action of the total operator reduces to the addition of a phase $2\phi$ to each photonic mode when the atom is in the excited state.

\noindent Without making any assumption about the value of $\phi$, the final density matrix of the system can be written as
\begin{equation}
    \begin{split}
         \rho^{(f)}_{sys}&= \hat{U}_{QM}^{2}\, \rho^{(i)}_{sys}\,\hat{U}_{QM}^{\dagger 2} \\ 
         &= \pi_{gg} 
    \otimes \rho'
    \otimes \ketbra{\varphi}{\varphi}
    +
    \pi_{ge}
    \otimes
   \rho' e^{ 2i \phi \hat{n}_0 } 
    \otimes \ketbra{\varphi}{\varphi} e^{  2i \phi \hat{n}_1 }\\ 
    &+
    \pi_{eg}  \otimes 
    e^{ -2i  \phi \hat{n}_0 } \rho'
    \otimes 
    e^{ -2i  \phi \hat{n}_1 } \ketbra{\varphi}{\varphi} +
    \pi_{ee} \otimes  
    e^{ -2i  \phi \hat{n}_0 } \rho' e^{2i  \phi \hat{n}_0 } 
    \otimes 
    e^{ -2i \phi \hat{n}_1 }  \ketbra{\varphi}{\varphi} e^{2i  \phi \hat{n}_1} 
    \end{split}
\end{equation}
To obtain the probabilities, we measure the control atom. The diagonal contributions in the atomic basis yield unity due to the normalization of the states, whereas nontrivial contributions arise from the off-diagonal atomic terms. Imposing the vacuum condition for the second mode, \(\ket{\varphi} = \ket{0}\), the probabilities reduce to
\begin{equation}
    p_{\pm} = \frac{1}{2} \left( 1 \pm \text{Tr}[\rho' e^{-2i \phi \hat{n}}] \pm \text{Tr}[e^{2i \phi \hat{n}} \rho'] \right). \label{eq:probb2QMWF}
\end{equation}
For the specific choice \(\phi = \pi/2\), the photon-number parity operator \(\hat{\Pi} = e^{i\pi \hat{n}}\) naturally emerges. So, the probabilities take the simple form
\begin{equation}
    p_{\pm} = \frac{1}{2} \left(1 \pm \text{Tr} [\hat{\Pi}\, \rho' ] \right).
\end{equation}
Finally, writing the state as \(\rho' = \hat{D}(\alpha)\rho\hat{D}^\dagger(\alpha)\) and using the definition of the Wigner function introduced in Ref.~\cite{Banaszek_1996}, we obtain
\begin{equation}
    W(\alpha) =\frac{2}{\pi}\big(\,p_+-p_-\big)=\frac{2}{\pi}\left<\hat{\sigma}_x\right>.
\end{equation}
This establishes a direct connection between the atomic measurement probabilities and the Wigner function of the field.

\section{Relative phase}

In Eq.~\eqref{eq: tr}, we establish the relation between the transmission and reflection coefficients for the beam splitter. Consequently, the two subspaces introduced in Eq.~\eqref{eq:UQM2}, corresponding to the ground ($g$) and excited ($e$) states, can be described by this type of beam splitter. In each subspace, the coefficients satisfy the relation
\begin{equation}
t_k = r_k + e^{i\phi_k},
\end{equation}
with $k = e, g$. From this expression, explicit forms of the coefficients in each subspace can be obtained. In particular, since $|r_g|\rightarrow 0$ when the atom is in the ground state and $|t_e|\rightarrow 0$ when the atom is in the excited state, we obtain
\begin{eqnarray}
     t_g &=  e^{i \phi_g} \,\,\,\, (\text{for } \ket{g}),\\
        r_e &=- e^{i \phi_e}\,\,\,\, (\text{for }   \ket{e}).
\end{eqnarray}
The action of these coefficients in each subspace leads to the following transformations of the field operators:
\begin{eqnarray}
    &(\hat{a}_0,\hat{a}_1)\overset{\hat{U}_g}{\longrightarrow} (\hat{a}_0 e^{i\phi_g},\hat{a}_1 e^{i\phi_g}),\\
    &(\hat{a}_0,\hat{a}_1)  \overset{\hat{U}_e}{\longrightarrow} (-\hat{a}_1 e^{i\phi_e},-\hat{a}_0 e^{i\phi_e}),
\end{eqnarray}
this allows us to write a more general version of the QM operator, which can be written as
\begin{equation}
    \hat{U}'_{QM}=\hat{\pi}_g \otimes \hat{U}_{g}+\hat{\pi}_e \otimes \hat{U}_e.
\end{equation}
Finally, the action of $\hat{U}_{QM}'$ on two incoming field modes prepared in arbitrary states $\ket{\psi}$ and $\ket{\varphi}$ and on the atomic state is given by
\begin{equation}
    \begin{split}
\hat{U}'_{QM}\ket{g}\ket{\psi}\ket{\varphi}&=\ket{g}\hat{U}_M^{(g)}\ket{\psi}\ket{\varphi}=\ket{g} e^{-i\phi_g\hat{n}_0}\ket{\psi}e^{-i\phi_g\hat{n}_1}\ket{\varphi}\\
&=(\mathbf{1}\otimes e^{-i\phi_g \hat{n}})\ket{g}\ket{\psi}\ket{\varphi}
    \end{split}
\end{equation}
\begin{equation}
    \begin{split}
\hat{U}'_{QM}\ket{e}\ket{\psi}\ket{\varphi}&=\ket{e}\hat{U}_M^{(e)}\ket{\psi}\ket{\varphi}=\ket{e} e^{-i\phi_e\hat{n}_0}\hat{\Pi}\ket{\varphi}e^{-i\phi_e\hat{n}_1}\hat{\Pi}\ket{\psi}\\
&=(\mathbf{1}\otimes e^{-i\phi_g \hat{n}})\ket{e} e^{-i(\phi_e-\phi_g)\hat{n}_0}\hat{\Pi}\ket{\varphi}e^{-i(\phi_e-\phi_g)\hat{n}_1}\hat{\Pi}\ket{\psi}
    \end{split}
\end{equation}
From these two expressions, we observe that the phase $\phi_g$ appears only as a global phase factor and is therefore physically irrelevant. Consequently, the physically relevant phase shift appearing in Eq.~\eqref{eq:QMket} can be identified as the phase difference between the two internal states of the QM,
\begin{equation}
    \phi = \phi_e - \phi_g .
\end{equation}

\end{document}